\documentclass[prd,twocolumn,superscriptaddress,showpacs,preprintnumbers,amsmath,amssymb]{revtex4}

\usepackage{graphicx}
\usepackage{dcolumn}
\usepackage{bm}

\begin{document}

\preprint{}

\title{Imprint of Explosion Mechanism on Supernova Relic Neutrinos}

\author{Ken'ichiro Nakazato}
 \email{nakazato@rs.tus.ac.jp}
 \affiliation{Department of Physics, Faculty of Science \& Technology, Tokyo University of Science, 2641 Yamazaki, Noda, Chiba 278-8510, Japan
}%

\date{\today}

\begin{abstract}
The spectrum and event rate of supernova relic neutrinos are calculated taking into account the dependence on the time it takes for the shock wave in supernova cores to revive. The shock revival time should depend on the still unknown explosion mechanism of collapse-driven supernovae. The contribution of black-hole-forming failed supernovae is also considered. The total event rate is higher for models with a longer shock revival time and/or a failed-supernova contribution. The hardness of the spectrum does not strongly depend on the shock revival time, but the spectrum becomes hard owing to the failed supernovae. Therefore, the shock-revival-time dependence of supernova relic neutrinos has different systematics from the fractions of failed supernovae.
\end{abstract}


\maketitle

\section{Introduction} \label{intro}
Supernova explosions are fundamental to the evolution of the universe and one of the central issues in astrophysics. Unfortunately, the explosion mechanism of collapse-driven supernovae is still an open question after a half century in spite of significant and long-lasting research efforts \cite{col66,sato75,bethe85,totani98}. However, progress in this field has been slow but steady \cite{thiel11,kotake12,janka12,burr13}. We now know that the core collapse of massive ($\gtrsim 10 M_\odot$) stars is bounced by the nuclear repulsion force and a shock wave is launched outward. It is difficult to reproduce such an explosion numerically because the shock wave tends to stall. The physics underlying shock propagation is still under debate.

We cannot study a collapsing core, which is embedded in a stellar envelope, through optical observations. Recently, Belczynski and co-workers \cite{bel12,fryer12} proposed a new method of estimating the time taken for shock propagation using the mass distributions of neutron stars and black holes. They concluded that a model in which the stalled shock revives within 100-200~ms of the bounce convincingly accounts for the lack of observed compact remnants in the mass range 2-$5M_\odot$. On the other hand, according to a recent numerical study \cite{yama13}, the shock is relaunched 300-400~ms after the bounce so as to produce the appropriate explosion energy and nickel yields. This shock revival time is important because it should depend on the explosion mechanism.

The shock revival time is also reflected in supernova relic neutrinos. Collapse-driven supernovae emit a large amount of MeV neutrinos, which constitute relic background radiation \cite{totani96,ando04,beacom10}. Thanks to the extraordinary efforts to reduce the background, the upper limit from the Super-Kamiokande detector \cite{malek03,bays12} is now close to the standard predictions \cite{hori09}. We will probably observe the signal of supernova relic neutrinos using future large detectors of Mton mass. Although theoretical predictions are important for analyses of the experimental data, the uncertainty of shock revival time was not investigated so far.

\begin{figure}
\begin{center}
\includegraphics[scale=0.84]{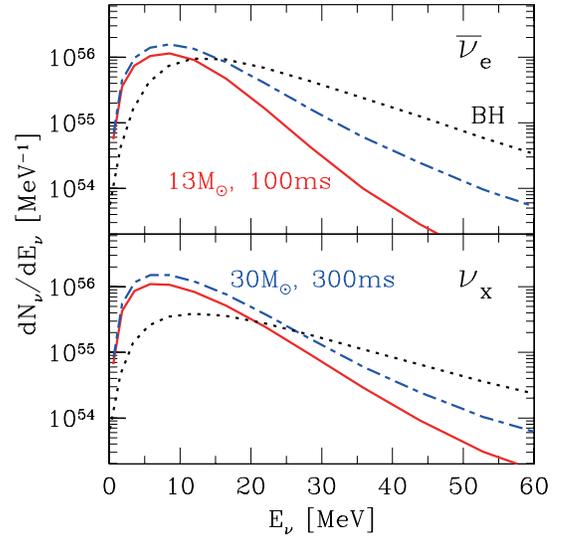}
\caption{Neutrino number spectra of supernovae with $13M_\odot$, $Z = 0.02$ and shock revival time of 100~ms (solid) and $30M_\odot$, $Z = 0.02$ and 300~ms (dot-dashed). The dotted lines denote the spectra of black-hole-forming failed supernova ($30M_\odot$ and $Z = 0.004$). The upper and lower panels corresponds to $\bar\nu_e$ and $\nu_x$ ($=\nu_\mu=\bar \nu_\mu=\nu_\tau=\bar \nu_\tau$), respectively.}
\label{snndb}
\end{center}
\end{figure}
In this paper, we evaluate the flux of supernova relic neutrinos with different shock revival times. We utilize the Supernova Neutrino Database \cite{self13}, where the neutrino light curves and spectra until 20~s are given for a variety of progenitor stellar masses (13-$50M_\odot$) and metallicities ($Z = 0.02$ and 0.004). In this data set, assuming spherical symmetry, the results of neutrino-radiation hydrodynamic simulations for the early phase and quasi-static evolutionary calculations of neutrino diffusion for the late phase are combined assuming shock revival at either 100, 200 or 300~ms after the bounce. The total emission number of supernova neutrinos increases with the shock revival time because of more material accreting to the collapsed core. It also increases with the core mass of progenitors, but is not monotonically related to the initial mass of progenitors due to the mass loss during the pre-collapse stages. In fact, the core mass of the model with the mass of $30M_\odot$ is the highest. The two extreme cases of the supernova neutrino number spectra are shown in Figure~\ref{snndb}. A supernova neutrino model with the similar time profile is shown in Ref.~\cite{fischer12}.

Cosmic sources of MeV neutrinos are not only ordinary collapse-driven supernovae but also failed supernovae, which collapse to a black hole without explosion \cite{sumi06,self08,fish09}. They are small in number but contribute to the overall flux because the luminosity and mean energy of their neutrinos are higher than those of ordinary collapse-driven supernovae \cite{luna09,lien10,yuk12}. In the Supernova Neutrino Database \cite{self13}, the progenitor model with $30M_\odot$ and $Z = 0.004$ corresponds to a failed supernova (Figure~\ref{snndb}). Below, we also consider the contribution due to failed supernovae.

\section{Setups} \label{setups}
Since the core collapse of some progenitors may result not in collapse-driven supernovae but in failed supernovae, we start with the total core-collapse rate, which is written as $R_{\rm CC} (z) = \zeta_{\rm CC} \dot{\rho}_\ast (z)$, as a function of the redshift $z$ with the cosmic star formation rate history $\dot{\rho}_\ast (z)$. Here, the conversion coefficient $\zeta_{\rm CC}$ is related to the initial mass function $\psi(M)$ as
\begin{equation}
\zeta_{\rm CC} = \frac{\int^{M_{\rm max}}_{M_{\rm min}} \psi (M) {\rm d}M}{\int^{100M_\odot}_{0.1M_\odot} M\psi (M) {\rm d}M},
\label{convcoef}
\end{equation}
where $M_{\rm max}$ and $M_{\rm min}$ are the maximum and minimum masses of progenitors that end with a core collapse, respectively. In this study, we set $M_{\rm min}=10M_\odot$ and $M_{\rm max}=100M_\odot$ for consistency with the progenitor models adopted in Ref.~\cite{self13}. We use the Salpeter A initial mass function \cite{bald03,hori11}, which scales as
\begin{equation}
\psi (M) \propto \left\{
\begin{array}{ll}
M^{-2.35}, & M \ge 0.5 M_\odot, \\
M^{-1.5}, & M < 0.5 M_\odot,
\end{array}
\right.
\label{salaimf}
\end{equation}
yielding $\zeta_{\rm CC}=0.0071 /M_\odot$. This value is close to that in Ref.~\cite{yuk12}.

For the star formation rate, we assume a smoothed broken power law of the form \cite{hopk06,yuk08,yuk12}
\begin{equation}
\dot{\rho}_\ast (z) = \dot{\rho}_0 \left[ (1+z)^{\alpha \eta} + \left( \frac{1+z}{B} \right)^{\beta \eta} + \left( \frac{1+z}{C} \right)^{\gamma \eta} \right]^{1/\eta},
\label{yuksfr}
\end{equation}
with $\alpha = 3.4$, $\beta = -0.3$, $\gamma = -3.5$ and $\eta = -10$. The coefficients $B=(1+z_1)^{1-\alpha/\beta}$ and $C=(1+z_1)^{(\beta-\alpha)/\gamma}(1+z_2)^{1-\beta/\gamma}$ make breaks at $z_1=1$ and $z_2=4$, respectively. We adopt $\dot{\rho}_0 = 0.02M_\odot {\rm yr}^{-1} {\rm Mpc}^{-3}$ for the cosmic star formation rate at $z=0$. The resultant core-collapse rate at $z=0$, $R_{\rm CC} (0) = 1.4 \times 10^{-4} {\rm yr}^{-1} {\rm Mpc}^{-3}$, in this model is consistent with the recently estimated nearby supernova rate within a distance of 6-15~Mpc of $1.5^{+0.4}_{-0.3} \times 10^{-4} {\rm yr}^{-1} {\rm Mpc}^{-3}$ \cite{mat12}; however see also Ref.~\cite{li11}. 

The fraction of core collapses that result in failed supernovae, $\varepsilon(z)$, is the most uncertain factor. According to stellar evolution theory, the mass loss is inefficient and the core is massive for metal-poor stars, as in the model adopted in Ref.~\cite{self13}. Therefore, we can surmise that $\varepsilon(z)$ is larger in a high-redshift universe. Here, we follow the idea of Y{\"u}ksel and Kistler \cite{yuk12}, who drew guidance from the rate of bright gamma-ray bursts, which evolves with $z$ more strongly than the star formation rate. We assume that $\varepsilon(z)$ grows with $z$ similarly and write $\varepsilon(z)=\varepsilon_0(1+z)^\delta$ \cite{kist09} with $\delta=1$ \cite{yuk12}. In this study, we examine two extreme cases, $\varepsilon_0=0$ and 0.1, for the fraction at $z=0$. Finally, we obtain the rates of collapse-driven supernovae $R_{\rm SN} (z) = (1-\varepsilon(z))R_{\rm CC} (z)$ and black-hole-forming failed supernovae $R_{\rm BH} (z) = \varepsilon(z)R_{\rm CC} (z)$.

The flux of supernova relic neutrinos on Earth is written as
\begin{eqnarray}
\frac{{\rm d}F(E_\nu)}{{\rm d}E_\nu} & = & c \int_0^{z_{\rm max}} \frac{{\rm d}z}{H_0 \sqrt{\Omega_m (1+z)^3 + \Omega_\Lambda}} \times \nonumber \\
 & & \Biggl[ R_{\rm SN} (z) \int^{M_{\rm max}}_{M_{\rm min}} \psi (M) \frac{{\rm d}N_{\rm SN}(M, E^\prime_\nu)}{{\rm d}E^\prime_\nu} {\rm d}M \nonumber \\
 & & {}+ R_{\rm BH} (z) \frac{{\rm d}N_{\rm BH}(E^\prime_\nu)}{{\rm d}E^\prime_\nu} \Biggr]
\label{yuksfr}
\end{eqnarray}
with the velocity of light $c$ and cosmological constants $H_0=70~{\rm km}/{\rm s}/{\rm Mpc}$, $\Omega_m=0.3$ and $\Omega_\Lambda=0.7$. The neutrino energy on Earth, $E_\nu$, is related to that at the redshift $z$, $E^\prime_\nu$, by $E^\prime_\nu=(1+z)E_\nu$. The initial mass function is normalized as $\int^{M_{\rm max}}_{M_{\rm min}} \psi (M) {\rm d}M=1$. We adopt the models of solar metallicity ($Z=0.02$) in the Supernova Neutrino Database \cite{self13} for the neutrino number spectrum of ordinary supernovae with mass $M$, ${\rm d}N_{\rm SN}(M, E^\prime_\nu)/{\rm d}E^\prime_\nu$. For the neutrino number spectrum of failed supernovae, ${\rm d}N_{\rm BH}(E^\prime_\nu)/{\rm d}E^\prime_\nu$, we use the model of $30M_\odot$ and $Z=0.004$ in the Supernova Neutrino Database \cite{self13} as a representative model.

The most promising channel for detecting supernova relic neutrinos is the inverse $\beta$ decay reaction of electron antineutrinos, $\bar{\nu}_e + p \to e^+ + n$. Note that the neutrinos undergo flavor conversion before detection. Recently, the mixing angle of neutrino oscillation $\theta_{13}$ has been confirmed to be nonzero \cite{t2k11,minos11} and evaluated to be $\sin^2 2\theta_{13} \sim 0.1$ \cite{dayab12,reno12,dchooz12}. Thus, at present, the most undetermined parameter in neutrino oscillation is the mass hierarchy. In this study, we consider the Mikheev-Smirnov-Wolfenstein effect expected in the stellar envelope \cite{wolf78,mik85} and assume the survival probability of $\bar{\nu}_e$ to be $\bar{P}_{ee}=0.68$ for the normal mass hierarchy and $\bar{P}_{ee}=0$ for the inverted mass hierarchy \cite{kotake06}. We do not take into account the neutrino-neutrino collective effects \cite{chakr11,dasgu12}, which are estimated to contribute at the 5-10\% level \cite{luna12}. We also neglect the effect of shock wave propagation on neutrino oscillation \cite{gala10}, which should be minor for a time-integrated signal \cite{kawa10}.

The positron spectrum due to supernova relic neutrinos is obtained as
\begin{equation}
\frac{{\rm d}N_{e^+}(E_{e^+})}{{\rm d}E_{e^+}} = N_t \sigma(E_{\bar{\nu}_e}) \frac{{\rm d}F(E_{\bar{\nu}_e})}{{\rm d}E_{\bar{\nu}_e}},
\label{convcoef}
\end{equation}
where $\sigma(E_{\bar{\nu}_e})$ is the cross section for the inverse $\beta$ decay \cite{stru03}. The positron energy is given as $E_{e^+}=E_{\bar{\nu}_e}-\Delta c^2$ with the neutron-proton mass difference $\Delta$. We set the number of target protons as $N_t=1.5 \times 10^{33}$ for Super-Kamiokande with a 22.5~kton fiducial volume and $N_t=3.7 \times 10^{34}$ for a future 560~kton detector such as Hyper-Kamiokande.

\section{Results and discussion} \label{resdis}
\begin{figure}
\begin{center}
\includegraphics[scale=0.84]{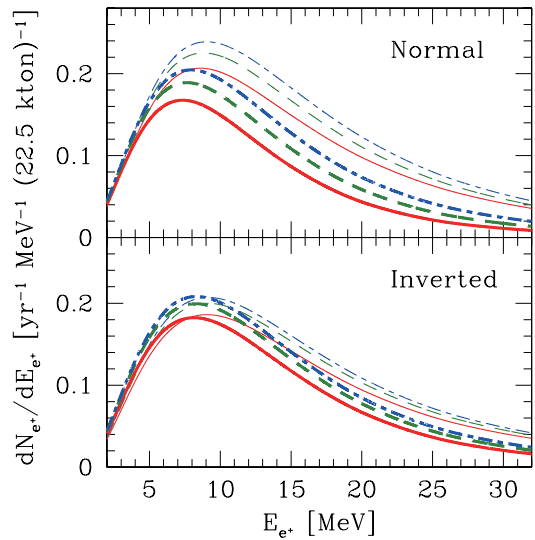}
\caption{Positron spectra in Super-Kamiokande over 1 year obtained using our supernova relic neutrino models. Solid, dashed and dot-dashed lines correspond to shock revival times of 100, 200 and 300~ms, respectively. Thick lines denote the case without failed supernovae ($\varepsilon_0=0$) and thin lines denote the case with failed supernovae ($\varepsilon_0=0.1$). The upper and lower panels show the spectra for normal and inverted mass hierarchies, respectively.}
\label{spectra}
\end{center}
\end{figure}

In Figure~\ref{spectra}, we show the positron spectra evaluated for Super-Kamiokande over 1 year obtained from our models with different shock revival times and fractions of failed supernovae. The event rate becomes higher with the inclusion of failed supernovae, as already known \cite{luna09,lien10,yuk12}. Furthermore, it also depends on the shock revival time: the models with a longer shock revival time have a higher event rate. This is for the following reason. The accretion of matter onto the bounced core continues until the shock revival. The released gravitational potential of the accreted matter is converted to the emitted neutrino energy. Thus, if the shock revives after a longer time, a larger amount of matter is accreted and more neutrinos are emitted.

Comparing the cases of normal and inverted mass hierarchies, the contribution of failed supernovae is clearer for the normal mass hierarchy. From the accretion of matter, $\nu_e$ and $\bar\nu_e$ are emitted more abundantly than $\nu_x$ ($=\nu_\mu=\bar \nu_\mu=\nu_\tau=\bar \nu_\tau$) owing to the capture of electrons and positrons on nucleons. In contrast, neutrinos of all species are emitted equivalently from the cooling of a proto-neutron star. Thus, all flavors have similar spectra for ordinary collapse-driven supernovae. On the other hand, for failed supernovae, neutrinos from the accretion of matter dominate and the total emission energy of $\bar\nu_e$ is about twice those of $\bar \nu_\mu$ and $\bar \nu_\tau$. The survival probability of $\bar{\nu}_e$ is 0.68 for the normal mass hierarchy, whereas all $\bar{\nu}_e$ convert to $\bar \nu_\mu$ or $\bar \nu_\tau$ for the inverted mass hierarchy. Therefore, the expected flux is larger for the normal mass hierarchy.

Since the mean energy of neutrinos emitted from failed supernovae is higher than those of ordinary collapse-driven supernovae, the spectrum of relic neutrinos becomes hard owing to the failed supernovae. On the other hand, the flux is larger but the spectrum is not too hard for the models with a longer shock revival time. This is because, after the shock revival, a proto-neutron star is not heated and the mean energy of neutrinos gradually decreases.

\begin{figure}
\begin{center}
\includegraphics[scale=0.84]{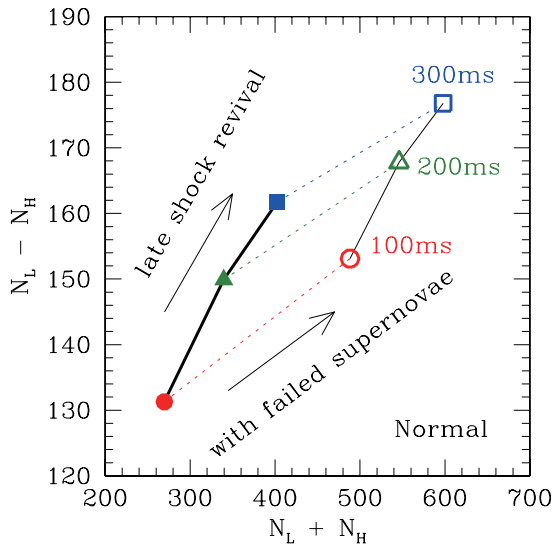}
\caption{Plots of $N_L+N_H$ versus $N_L-N_H$ for the normal mass hierarchy, where $N_L$ and $N_H$ are the event rates per 560~${\rm kton} \times 10$~year in the positron energy ranges of 10~${\rm MeV} \leq E_{e^+} \leq 18$~MeV and 18~${\rm MeV} \leq E_{e^+} \leq 26$~MeV, respectively. Filled symbols represent the case without failed supernovae ($\varepsilon_0=0$) and empty symbols represent the case with failed supernovae ($\varepsilon_0=0.1$). Circles, triangles and squares represent the shock revival times of 100, 200 and 300~ms, respectively. Lines are shown as a guide to the eyes.}
\label{fignor}
\end{center}
\end{figure}
\begin{figure}
\begin{center}
\includegraphics[scale=0.84]{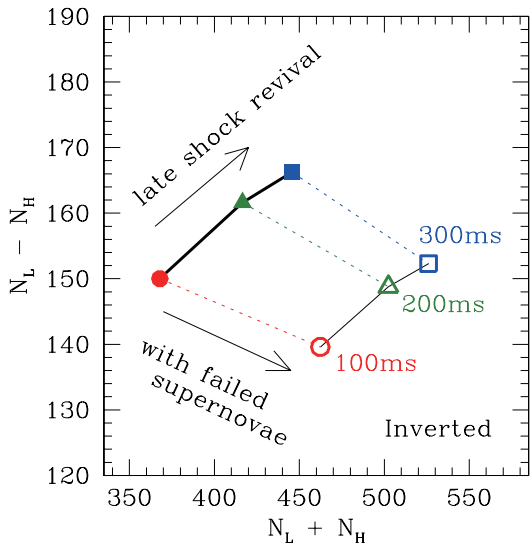}
\caption{Same as Figure~\ref{fignor} but for the inverted mass hierarchy.}
\label{figinv}
\end{center}
\end{figure}
Here, we consider the event rates in the positron energy ranges of 10~${\rm MeV} \leq E_{e^+} \leq 18$~MeV, $N_L$, and 18~${\rm MeV} \leq E_{e^+} \leq 26$~MeV, $N_H$. For $N_H$, the upper limit of the supernova relic neutrinos has already been given by Super-Kamiokande \cite{malek03}. By improving the capabilities of detectors tagging the inverse $\beta$ decay with Gd \cite{beacom04}, the relic neutrino signal should be detectable for $N_L$. The sum of $N_L$ and $N_H$, $N_L+N_H$, represents the total event rate, and the spectral features are reflected in the difference between $N_L$ and $N_H$, $N_L-N_H$. 

Plots of $N_L+N_H$ versus $N_L-N_H$ expected using a 560 kton detector over 10 years for the normal and inverted mass hierarchies are shown in Figures~\ref{fignor} and \ref{figinv}, respectively. The total event rate is 250-600 for all cases and it is higher for the models with a longer shock revival time and/or a failed-supernova contribution. On the other hand, the shock-revival-time dependence and failed-supernova contribution exhibit different trends on the $N_L+N_H$ versus $N_L-N_H$ plane. Since the relic neutrino spectrum becomes hard owing to failed supernovae, the event rate of high-energy neutrinos ($N_H$) becomes much higher. In contrast, $N_H$ and $N_L$ increase equally if the shock revival is retarded. Therefore, for the same total event rate ($N_L+N_H$), the cases with a longer shock revival time have a larger value of $N_L-N_H$ than the cases with failed supernovae. This feature is clearer for the case of the inverted mass hierarchy.

Since the shock revival time should depend on the still unknown explosion mechanism of collapse-driven supernovae \cite{bel12,fryer12,self13}, supernova relic neutrinos would reflect the physics underlying the explosion. The mass hierarchy would be determined experimentally \cite{winter13} and the failed supernova fraction, $\varepsilon(z)$, may be estimated as the difference between the measured cosmic supernova rate and the total core-collapse rate predicted from the star formation rate \cite{hori11}. However, from observational point of view, since there is the non-negligible atmospheric background \cite{fogli05}, careful discussion taking into account statistical and systematical errors is mandatory.

Since this is the first attempt to evaluate the spectrum and event rate of supernova relic neutrinos taking into account the dependence on the shock revival time, there are some issues beyond the scope of this study. Here, we have assumed that all supernovae have the same shock revival time, whereas it may depend on the progenitor mass and/or metallicity. The supernova relic neutrino flux reflects an averaged shock revival time and, therefore, observations of neutrinos from a single Galactic supernova and supernova relic neutrinos are complementary. The collective oscillation and shock propagation also affect the supernova relic neutrino flux \cite{lund13} while the qualitative features described in this paper should be unchanged. Although the neutrino signal from a failed supernova is sensitive to the nuclear equation of state \cite{sumi06}, we have used a single model by Shen~et~al. \cite{shen98} in this study. Future nuclear experiments such as heavy-ion collisions are important to fix the ambiguity.

\section{Summary} \label{summary}
In summary, we have investigated the relic background radiation from collapse-driven and failed supernovae. This study is the first to provide the dependence on the shock revival time, which reflects the unknown explosion mechanism. We have found that the relic neutrino flux is larger for models with a longer shock revival time and/or a failed-supernova contribution. The hardness of the spectrum does not strongly depend on the shock revival time, whereas the spectrum becomes hard owing to failed supernovae. Furthermore, we have found that the shock-revival-time dependence and failed-supernova contribution exhibit different trends on the $N_L+N_H$ versus $N_L-N_H$ plane. We hope that our conclusion will be shown to be valid by the future progress in astronomical observations and nuclear experiments.

\begin{acknowledgments}
The author is grateful to H. Suzuki for fruitful discussions and appreciates the cooperation of D. Kato received in the early stage of this study. This work was partially supported by Grants-in-Aids for the Scientific Research (Nos.~20105004, 23840038, 24105008) from MEXT in Japan.
\end{acknowledgments}

\bibliographystyle{apsrev}
\bibliography{apssamp}

\end{document}